\documentclass[fleqn,usenatbib]{mnras}

\usepackage{newtxtext,newtxmath}
\usepackage[T1]{fontenc}

\DeclareRobustCommand{\VAN}[3]{#2}
\let\VANthebibliography\thebibliography
\def\thebibliography{\DeclareRobustCommand{\VAN}[3]{##3}\VANthebibliography}

\usepackage{graphicx}	% Including figure files
\usepackage{amsmath}	% Advanced maths commands

\title[]{Rescued from oblivion: detailed analysis of archival {\it Spitzer} data of SN~1993J}

\author[Sz. Zsíros et al.]{
Szanna Zsíros,$^{1}$\thanks{email: \href{mailto:szannazsiros@titan.physx.u-szeged.hu}{szannazsiros@titan.physx.u-szeged.hu}}
Andrea P. Nagy,$^{1,2}$
and Tamás Szalai$^{1,2,3}$
\\
$^{1}$Department of Optics and Quantum Electronics, Institute of Physics, University of Szeged, H-6720 Szeged, D\'om t\'er 9, Hungary\\
$^{2}$Konkoly Observatory, Research Centre for Astronomy and Earth Sciences, H-1121 Budapest, Konkoly Thege Mikl\'os \'ut 15-17, Hungary\\
$^{3}$MTA-ELTE Lend{\"u}let Milky Way Research Group, Hungary
}

\date{Accepted XXX. Received YYY; in original form ZZZ}

\pubyear{2021}

\begin{document}
\label{firstpage}
\pagerange{\pageref{firstpage}--\pageref{lastpage}}
\maketitle

\begin{abstract}
We present an extensive analysis of the late-time mid-infrared (mid-IR) evolution of Type IIb SN~1993J from 10 up to 26 years post-explosion based on archival -- mostly previously unpublished --  photometric data of Spitzer Space Telescope in conjunction with an archival IRS spectrum. SN~1993J is one of the best-studied supernovae (SNe) with an extensive, decade-long multi-wavelength dataset published in various papers; however, its detailed late-time mid-IR analysis is still missing from the literature. Mid-IR data follows not just the continuously cooling SN ejecta but also late-time dust formation and circumstellar interaction processes. We provide evidence that the observed late-time mid-IR excess of SN~1993J can be described by the presence of two-component local dust with a dust mass of $\sim(3.5-6.0)\times 10^{-3} M_{\odot}$ in case of a partly silicate-based dust composition. Source of these components can be either newly-formed dust grains, or heating of pre-existing dust via ongoing CSM interaction detected also at other wavelengths. If it is newly-formed, dust is assumed to be located both in the unshocked inner ejecta and in the outer cold dense shell, just as found in the Cassiopeia A remnant and also assumed in other dust-forming SNe in a few years after explosion.
\end{abstract}


%%%%%%%%%%%%%%%%%%%%%%%%%%%%%%%%%%%%%%%%%%%%%%%%%%

%%%%%%%%%%%%%%%%% BODY OF PAPER %%%%%%%%%%%%%%%%%%

\section{Introduction}\label{sec:intro}

Supernovae (SNe) are cataclysmic endings of evolved massive stars or of white dwarfs located in binary systems. 
They are unique astrophysical laboratories for studying not only the extremely energetic final explosions, but also their impact on their environment, and the observable signs of pre-explosion stellar evolution processes.
Multi-channel follow-up of the evolution of the expanding SN ejecta and of its interaction with the ambient medium plays an important role in these researches.

The optical follow-up of SN explosions is one of the most plausible and widely applied research methods; nevertheless, data collected in other wavelength ranges allow us to get a deeper insight into the physical background of these events.
Mid-infrared (mid-IR) observations provide numerous advantages for following the late-time evolution of SNe due to the increased sensitivity to the expanding and cooling ejecta and the smaller impact by interstellar extinction.
This wavelength region also covers atomic and molecular emission lines generated by shocked, cooling gas \citep{reach06}. Moreover, mid-IR observations are also sensitive to warm dust either in the SN ejecta or in the pre-existing circumstellar matter (CSM).

Newly-condensed SN dust may form not only in the ejecta but also in a cool dense shell (CDS) across the contact discontinuity between the shocked CSM and shocked ejecta \citep[see e.g][]{chugai04,pozzo04,mattila08}. 
Besides, late-time mid-IR excess may also emerge from heated pre-existing dust grains. 
In the shocked CSM, heating can be collisional, while grains in the more distant, unshocked CSM are assumed to be radiatively heated by the peak SN luminosity or by energetic photons generated during CSM interaction, thereby forming an IR echo \citep[see e.g.][]{bode80,dwek83,graham86,sugerman03,kotak09,fox10}. In these cases, dust can be a useful probe of the CSM characteristics and of the pre-SN mass loss from either the progenitor or companion star \citep[see e.g.][for a review]{gall11}.

In the last two decades, one of the most important sources of mid-IR SN data was NASA's recently shutdown Spitzer Space Telescope (hereafter {\it{Spitzer}}), which provided valuable data during both its cryogenic (2003-2009) and post-cryogenic (2009-2020) missions. 
Beyond a few large-scale SN surveys -- e.g. SPIRITS project \citep[SPitzer InfraRed Intensive Transients Survey, a systematic study of nearby galaxies, see][]{tinyanont16,kasliwal17,jencson19}, or further ones focused on CSM-interacting (Type IIn) SNe \citep{fox11,fox13} --, and several single-object studies, many other objects may appear on non-targeted archival {\it Spitzer} images. \citet{szalai19,szalai21} presented the most complete analyses of mid-IR SN data, including $\sim$120 positively detected objects from $\sim$1100 SN sites imaged by {\it Spitzer}.

These latter studies focus primarily on the comprehensive examination of 3.6 \& 4.5 $\mu$m photometric dataset of SNe obtained with the InfraRed Array Camera (IRAC) detector of {\it Spitzer}.
Nevertheless, more extended {\it Spitzer} data of some other, nearby SNe are also available, including additional longer-wavelength measurements obtained with IRAC (5.8 \& 8.0 $\mu$m), as well as with Multiband Imaging Photometer (MIPS, 24 $\mu$m) and {with} InfraRed Spectrograph for Spitzer (IRS, $\sim$5$-$16 $\mu$m). Detailed works based on extended mid-IR datasets have been published e.g. for several Type II-P SNe \citep[e.g.][]{kotak09,fabbri11,meikle11,szalai11,szalai13}, but also for e.g. SNe~1978K \citep{tanaka12}, 1980K \citep{sugerman12}, and for the famous SN~1987A \citep{bouchet06,dwek10,arendt16,arendt20}.

Another nearby and famous object is SN~1993J, the prototype of Type IIb explosions \citep[forming a transitional group between H-rich Type II and H-free Type Ib/c core-collapse (CC)SNe, see e.g.][]{filippenko93,nomoto93}.
%SN~1993J was discovered in NGC 3031 (M81) on 1993 March 28.9 UT by an amateur astronomer, F. Garcia \citep[see in][]{ripero93}. 
Due to its proximity \citep[3.63$\pm$0.31Mpc,][]{freedman01}, and its fortunate location in the host galaxy, M81, SN~1993J has become one of the most well-observed SNe, possessing various long-term multi-wavelength dataset s and detailed analyses published in the literature. 

It is also worth highlighting that SN~1993J is one of the few SNe, whose progenitor star i) was directly identified on pre-explosion images -- found to be a K0-type red supergiant (RSG) star by \citet{aldering94,cohen95}, also confirmed by the observation of its disappearance on late-time images \citep{maund09} -- and ii) was found to be located in a massive binary system.
Regarding the latter point, models based on the early light curves (LCs) already predicted that the progenitor might lost most of its H-rich envelope as a result of its evolution in a close binary system \citep[see e.g.][]{podsiadlowski93,ray93,shigeyama94,woosley94}. 
As a further evidence, \citet{vandyk02} and \citet{maund04} showed that near-ultraviolet (NUV) and B-band excess flux may indicate the possible presence of a hotter (B2 type) putative companion of the exploded RSG star, which was confirmed later via the direct detection of the NUV continuum of the secondary component \citep[][]{fox14}. 

Moreover, SN~1993J has also showed various signs of ongoing (moderate-level) CSM interaction from early to late phases in radio and X-ray \citep[e.g.][]{vandyk94,zimmermann94,marcaide97,marcaide09,bietenholz01,bietenholz03,bartel02,weiler07,chandra09,mvidal11,dwarkadas14}, and also in the late-time optical spectra \citep[e.g.][]{matheson00,milisavljevic12,smith17}.
Signs of dust formation have been also published based on early-time near-IR \citep{matthews02}, and late-time optical spectral analysis \citep{bevan17}, as will be discussed in more detail later.

Thanks to SPIRITS and some further programs, M81 galaxy has been a recurrent target of {\it Spitzer} during its whole operation time, and SN~1993J has been detectable up to the final times.
Nevertheless, to date, only the 3.6 and 4.5 $\mu$m photometry of the object has been published by \citet{tinyanont16} \citep[and adopted by us in][]{szalai19}; detailed analysis of its extended mid-IR dataset (included also longer-wavelength IRAC, MIPS, and IRS data) is still missing from the literature.

In this paper, we present the full late-time mid-IR evolution of SN~1993J between $\sim10-26$ years after explosion, and a detailed analysis on the possible connection of the detected mid-IR radiation with assumed dust-formation and CSM-interaction processes.
In Section \ref{sec:obs}, we describe the steps of data reduction, while, in Section \ref{sec:res}, we present the results of our analysis based on modeling the mid-IR LCs and spectral energy distributions. Finally, we draw our conclusions in Section \ref{sec:concl}.

\section{Observations and data analysis}\label{sec:obs}

\subsection{Photometry on {\it{Spitzer}} images}

\begin{figure*}
        \includegraphics[width=\textwidth]{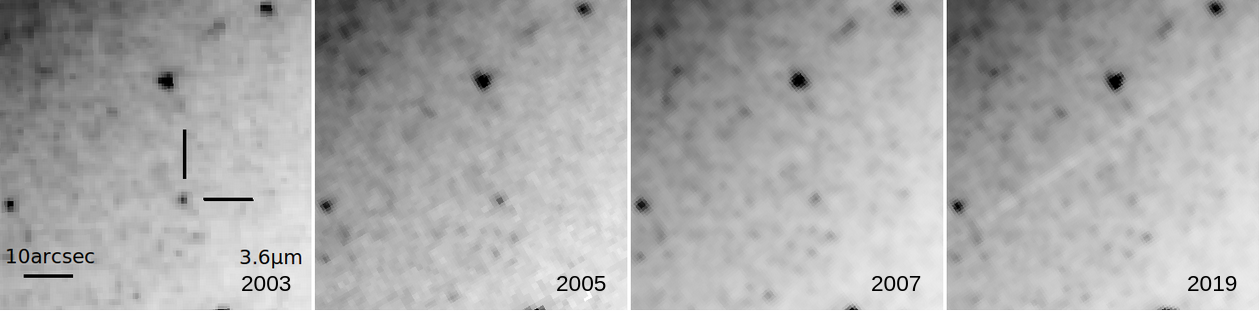}
        \includegraphics[width=\textwidth]{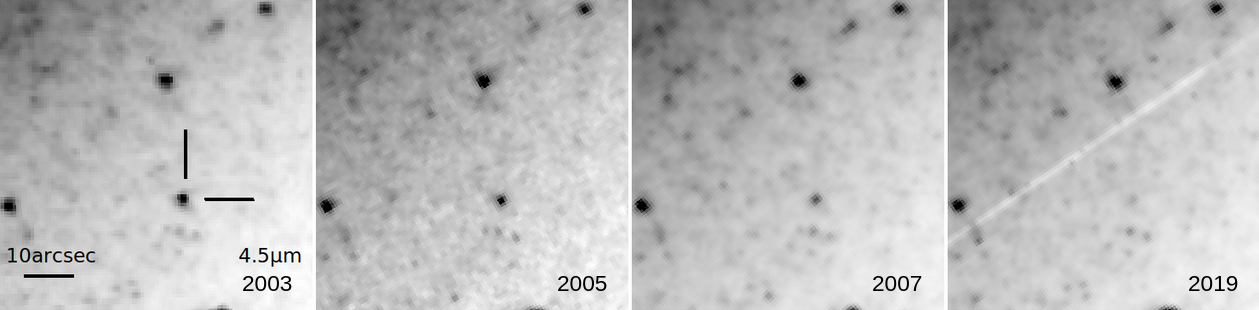}
        \includegraphics[width=\textwidth]{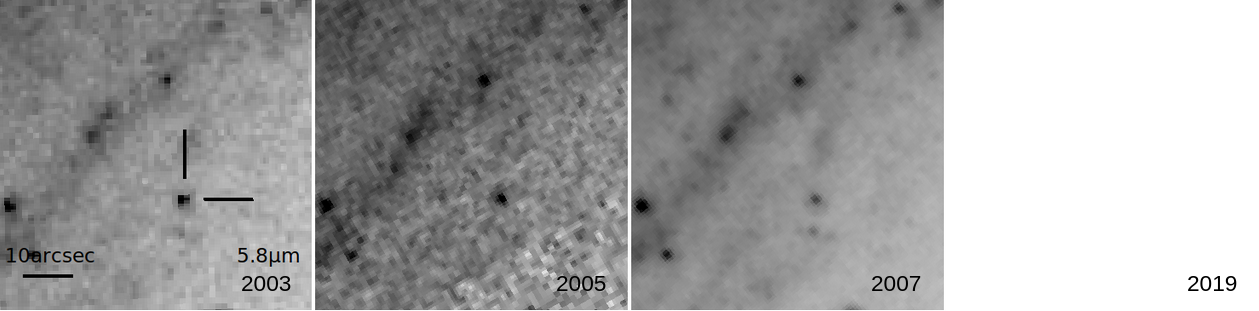}
        \includegraphics[width=\textwidth]{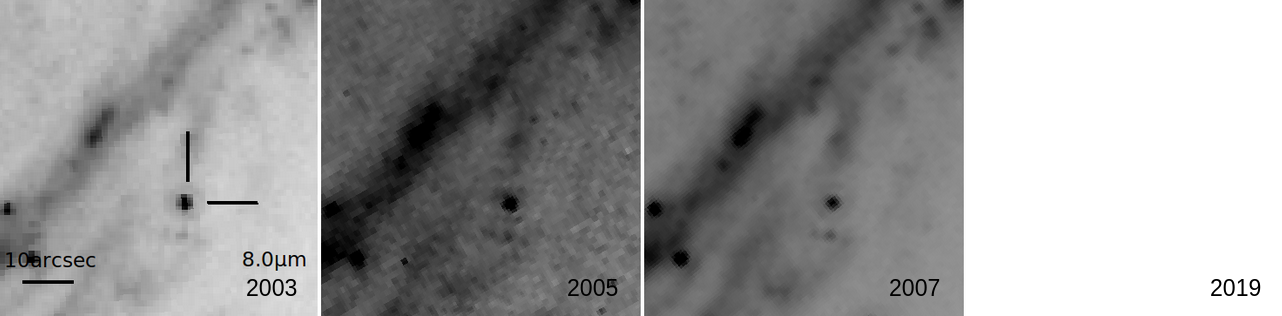}
        \includegraphics[width=\textwidth]{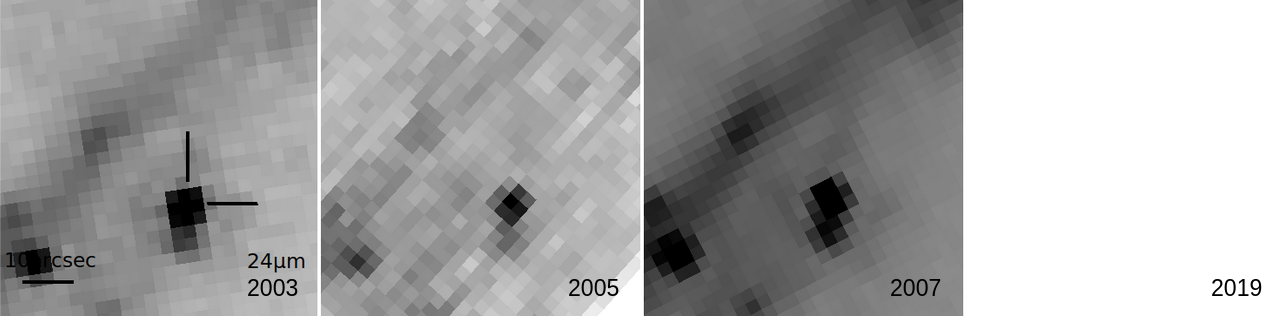}
    \caption{Snapshots from the late-time mid-IR evolution of SN~1993J from 2003, 2005, 2007 and 2019, showing {\it Spitzer}'s IRAC 3.6, 4.5, 5.8, 8.0, and MIPS 24.0 PBCD images. The position of the SN is marked on the first image for every channel. After 2009, only 3.6 and 4.5$\mu$m channels have been usable, thus, last epochs are missing from rows 3-5. }
    \label{fig:36}
\end{figure*}

In order to investigate the late-time mid-IR evolution of SN~1993J, we downloaded all of its archival IRAC and MIPS post-basic calibrated data (PBCD) from the Spitzer Heritage Archive\footnote{\href{https://sha.ipac.caltech.edu/applications/Spitzer/SHA/}{https://sha.ipac.caltech.edu/applications/Spitzer/SHA/}} (SHA). 
First, we checked the visibility of the SN on the images. At epochs between 2003 and 2009, a slowly fading point source is clearly seen at the position of SN~1993J at all IRAC channels as well as at MIPS 24 $\mu$m images (a few MIPS 70 and 160 $\mu$m images also cover this field, but the SN can not be identify on any of them, probably because of the limitation in both spatial and spectral resolution of these channels).
After 2009, SN~1993J is still detectable on IRAC 3.6 and 4.5 $\mu$m images, however, after $\sim$7500 days the object seems to fade into the local background. Fig. \ref{fig:36} illustrates the late-time mid-IR evolution of the SN on a set of IRAC/MIPS images.
While at a few epochs, some images do not cover the field of SN and we also had to eliminate a few more images because of pixel errors, we were able to carry out photometry on set of IRAC and MIPS images at 27 and 5 different epochs, respectively. 

Since, fortunately, SN~1993J is located in a less crowded region of the host galaxy, we were able to carry out simple aperture photometry on PBCD frames using the {\it phot} task of \textsc{iraf}\footnote{\href{http://iraf.noao.edu}{http://iraf.noao.edu}} (Image Reduction and Analysis Facility) software package. For the IRAC images, we used $3-3-7$ aperture configuration (given in native $\sim$1.2 \arcsec/pixel values), i.e. an aperture radius of 3.6\arcsec and a background annulus from 3.6\arcsec to 8.4\arcsec. 
For MIPS 24 $\mu$m images (image scale: 2.45\arcsec/pixels), we applied an aperture radius of 5\arcsec and a background annulus from 5\arcsec to 12\arcsec. 
The detectors of {\it Spitzer} uses fixed aperture sizes, hence to obtain the monochromatic fluxes of the SN, aperture corrections are needed. We used the values of 1.124, 1.270, 1.143, and 1.234 for the IRAC 3.6, 4.5, 5.8 and 8.0 $\mu$m channels, respectively, and 2.12 for the MIPS 24 $\mu$m channel (according to IRAC and MIPS instrumental handbooks). 

\begin{table*}
	\centering
	\caption{Mid-IR fluxes of SN~1993J determined from IRAC and MIPS data. We adopted the date of the explosion $t_0=$2 449 074.0 \citep{lewis94}. Some set of observations were carried out with a difference of only a few days; in these cases, we give the mean values of the measured fluxes. Program IDs: 1035 (PI Steven Willner), 717 (PI George Rieke), 1101 (PI Calibration, IRAC), 159 (PI Robert Kennicutt), 121 (PI Giovanni Fazio), 1860 (PI Calibration, MIPS), 40619 (PI Rubina Kotak), 80015 (PI Christopher S Kochanek) and 10136 (PI Mansi Kasliwal).}
	\label{tab:all_data}
	\begin{tabular}{lcccccc}
		\hline
		MJD & epoch & $F_{\nu,3.6}$ & $F_{\nu,4.5}$ & $F_{\nu,5.8}$ & $F_{\nu,8.0}$ & $F_{\nu,24.0}$\\
		(days) & (days) & ($\mu$Jy) & ($\mu$Jy) & ($\mu$Jy) & ($\mu$Jy) & ($\mu$Jy)\\
		\hline
52949 & 3875 & 131$\pm$ 22 & 234 $\pm$ 26 & 320 $\pm$ 31 & 1059 $\pm$ 59 & - \\
52967 & 3893 & - & - & - & - & 5078 $\pm$ 585\\
52978 & 3904 & 160 $\pm$ 23 & 230 $\pm$ 25 & 325 $\pm$ 30 & 924 $\pm$ 56 & -\\
53126 & 4052 & 133 $\pm$ 21 & 204 $\pm$ 24 & 281 $\pm$ 30 & 844 $\pm$ 54 & -\\
53294 & 4220 & - & - & - & - & 4510 $\pm$ 536\\
53296 & 4222 & - & - & - & - & 4882$\pm$ 558\\
53361 & 4287 & 133 $\pm$ 22 & 180 $\pm$ 23 & 206 $\pm$ 27 & 731 $\pm$ 51 & - \\
53472 & 4398 & - & - & - & - & 4130 $\pm$ 525\\
53497 & 4423 & 122 $\pm$ 21 & 169 $\pm$ 22 & 271 $\pm$ 32 & 653 $\pm$ 51 & - \\
53667 & 4593 & 97 $\pm$ 19 & 155 $\pm$ 21 & 233 $\pm$ 28 & 615 $\pm$ 48 & -\\
54419 & 5345 & 92 $\pm$ 19 & 106 $\pm$ 18 & 148 $\pm$ 22 & 384 $\pm$ 39 & - \\
54433 & 5359 & - & - & - & - & 3672 $\pm$ 484\\
55737 & 6663 & 83 $\pm$ 16 & 61 $\pm$ 14 & - & - & - \\
55939 & 6865 & 73 $\pm$ 16 & 68 $\pm$ 14 & - & - & - \\
56670 & 7596 & 63 $\pm$ 15 & 54 $\pm$ 13 & - & - & - \\
56699 & 7625 & 67 $\pm$ 16 & 52 $\pm$ 13 & - & - & - \\
56822 & 7748 & 71 $\pm$ 16 & 52 $\pm$ 13 & - & - & - \\
57061 & 7987 & 63 $\pm$ 15 & 49 $\pm$ 13 & - & - & - \\
57066 & 7992 & 67 $\pm$ 16 & 49 $\pm$ 13 & - & - & - \\
57081 & 8007 & 66 $\pm$ 16 & 46 $\pm$ 12 & - & - & - \\
57187 & 8113 & 73 $\pm$ 16 & 49 $\pm$ 12 & - & - & - \\
57398 & 8324 & 66 $\pm$ 16 & 45 $\pm$ 12 & - & - & - \\
57406 & 8332 & 61 $\pm$ 15 & 43 $\pm$ 12 & - & - & - \\
57426 & 8352 & 71 $\pm$ 17 & 45 $\pm$ 13 & - & - & - \\
57772 & 8698 & 64 $\pm$ 15 & 41 $\pm$ 12 & - & - & - \\
57928 & 8854 & 64 $\pm$ 15 & 47 $\pm$ 12 & - & - & - \\
58153 & 9079 & 62 $\pm$ 16 & 37 $\pm$ 11 & - & - & - \\
58289 & 9215 & 62 $\pm$ 15 & 35 $\pm$ 11 & - & - & - \\
58524 & 9450 & 68 $\pm$ 16 & 40 $\pm$ 12 & - & - & - \\
58574 & 9500 & 64 $\pm$ 15 & 45 $\pm$ 12 & - & - & - \\
58662 & 9588 & 60 $\pm$ 15 & 41 $\pm$ 12 & - & - & - \\
58716 & 9642 & 62 $\pm$ 15 & 34 $\pm$ 11 & - & - & - \\
		\hline
	\end{tabular}
\end{table*}

\begin{figure}
	\includegraphics[width=\columnwidth]{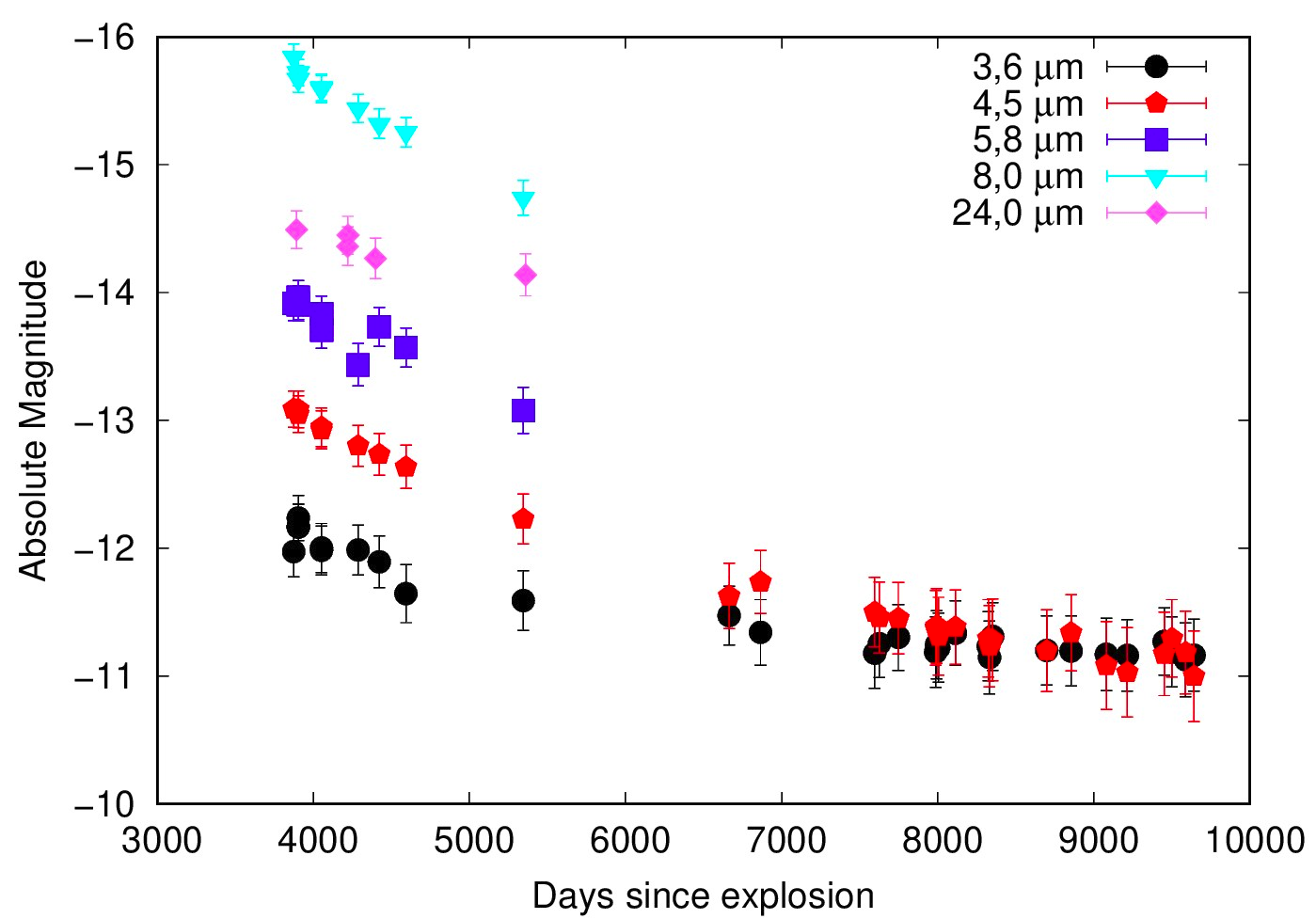}
    \caption{Mid-IR light curves of SN~1993J. After $\sim$7500 days, the object becomes close to the detection limit at both 3.6 and 4.5 $\mu$m channels.}
    \label{fig:lc}
\end{figure}

All measured mid-IR fluxes of SN~1993J are listed in Tab.~\ref{tab:all_data}.
We adopted the explosion date of $t_0$=2 449 074.0 MJD from \citet{lewis94}. Flux uncertainties are generally based on photon statistics provided by {\it phot}. In the cases of 3.6 and 4.5 $\mu$m data, we compared our photometric results to that of \citet{tinyanont16} and found a good agreement.
In Fig.~\ref{fig:lc}, we show the mid-IR LCs of SN~1993J in absolute (Vega) magnitudes. During these calculations, we adopted the distance modulus of $\mu_0$=27.80$\pm$0.08 mag \citep{freedman01}, and the total (Galactic + host) reddening value of E(B$-$V)=0.19$\pm$0.09 mag \citep{richardson06}. At $\sim$7500 days, both 3.6 $\mu$m and 4.5 $\mu$m LCs seem to reach a plateau, thus, in the later part of our analysis, we only used the {\it Spitzer} data of SN~1993J obtained before $\sim$7500 days.

\subsection{IRS spectrum}

A single, low-resolution, previously unpublished IRS spectrum of SN~1993J has been also obtained in 2008 (PID 40619, PI: R. Kotak; AORkey 23107072). We downloaded this spectrum from the database of Combined Atlas of Sources with Spitzer IRS Spectra \footnote{\href{https://cassis.sirtf.com/}{https://cassis.sirtf.com/}}\citep[CASSIS, ][]{lebouteiller11}. Since CASSIS identified the source as point-like and considered that Optimal extraction produces the best flux-calibrated spectrum, we followed this instruction during downloading.

\section{Results}\label{sec:res}

\subsection{Comparative analysis of mid-IR LC evolution of SN~1993J}\label{subsec:res_lc}

As has been previously showed by \citet{tinyanont16} and by \citet{szalai19}, SN~1993J is a unique case within Type IIb (and also within all 'normal' stripped-envelope CC) SNe with its very long ($\sim$3800$-$9600 days) mid-IR photometric coverage. While two other, nearby SNe IIb, 2011dh and 2013df, have been also followed with {\it Spitzer} during several years, both of them have become undetectable at either 3.6, or 4.5 $\mu$m channels at $\sim$1000 days after explosion. (Note that these are more distant objects than SN~1993J, located at $\sim$8 and $\sim$16 Mpc, respectively, and appear on more complex IR background on {\it Spitzer} images).

Because of lack of early-time {\it Spitzer} data of SN~1993J, these all would make the direct comparison of its mid-IR evolution with that of other SNe IIb impossible. At the same time, during the first $\sim$250 days, a nice set of ground-based near-IR photometry has been obtained on SN~1993J, as presented by \citet{matthews02}. These data include also L-band measurements, which can be matched with {\it Spitzer} 3.6 $\mu$m channel. While there is a large gap between $\sim$250 and $\sim$3800 days, data of SNe 2011dh and 2013df can help to make a bridge between early- and late-time mid-IR data of SN~1993J.

SN~2013df has an especially important role here. As it was shown in more studies, SNe 1993J and 2013df are spectroscopic twins regarding both of their early- and late-time properties \citep{mg14,vandyk14,maeda15,szalai16}, also resulting in the conclusion that both of them arose from a progenitor with an extended (several hundreds of $R_{\odot}$) H-envelope. Their optical LCs are also quite similar; however, based on the detailed analysis of both new and previously published data, \citet{szalai16} showed that SN~2013df was less luminous than 1993J (with a difference of $\sim$0.7 mag in every optical bands). While \citet[][based on optical spectra]{maeda15} and \citet[][based on radio and X-ray observations]{kamble16} revealed signs of moderate late-time CSM interaction in case of SN~2013df, very similar to that previously was found in 1993J, it can be assumed that they also have similar circumstellar environments.

\begin{figure*}
	\includegraphics[width=\textwidth]{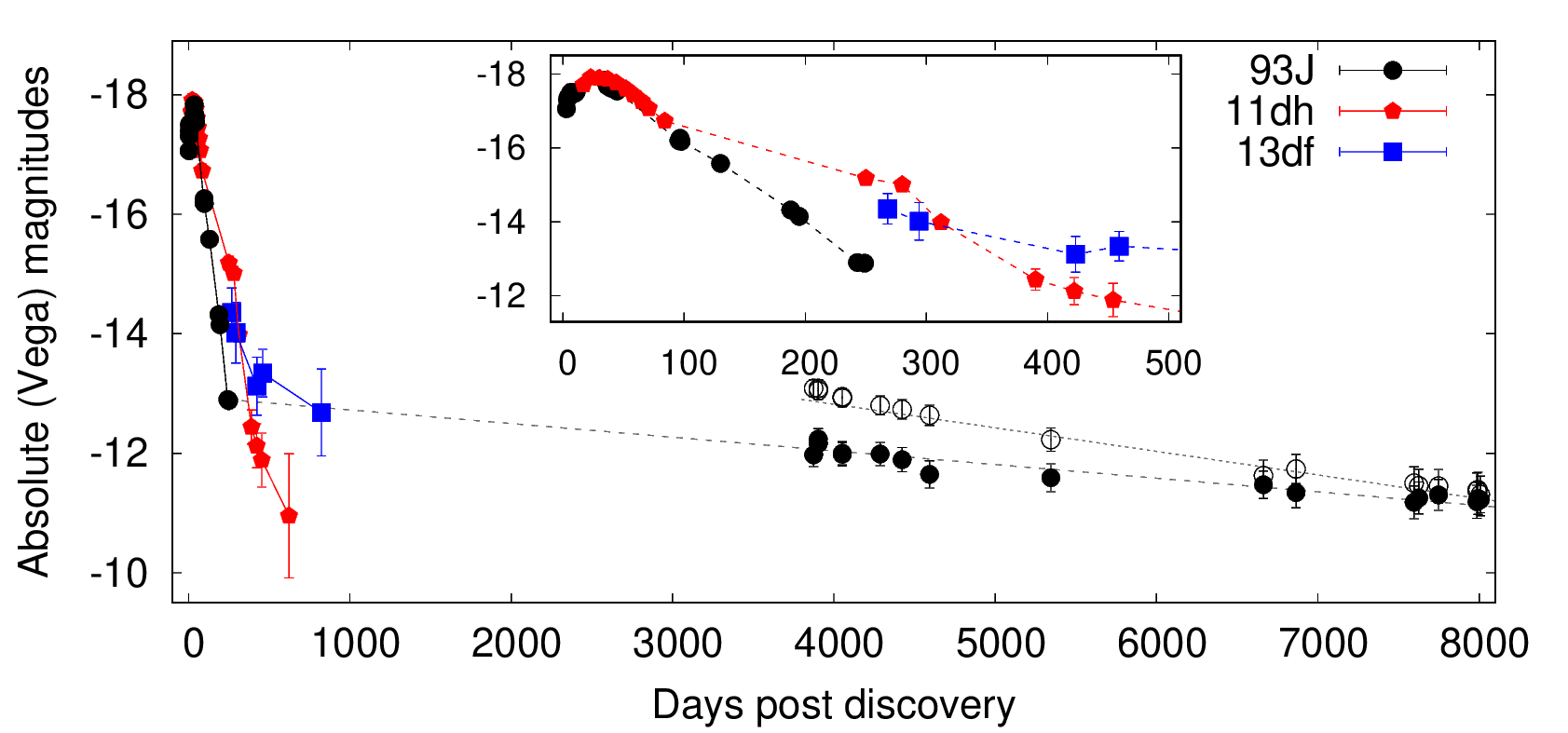}
    \caption{Long-term mid-IR LC evolution of SN~1993J \citep[early-time data are adopted from][]{matthews02} compared to that of SNe IIb 2011dh \citep{helou13,ergon15} and 2013df. Filled and empty symbols denote 3.6 and 4.5 $\mu$m data, respectively (early-time data of SN~1993J are from ground-based L-band measurements). The small insert shows the early-time 3.6$\mu$m/L-band data only. Dotted and dashed gray curves show the declining rates of SN~1993J mid-IR brightness. See text for details.}
    \label{fig:irlc_long}
\end{figure*}

In Fig. \ref{fig:irlc_long}, we compare long-term 3.6 $\mu$m LC evolution of SN~1993J to that of SNe~2013df \citep{tinyanont16,szalai16} and 2011dh \citep{helou13,ergon15}. Unfortunately, there are no overlapping data regarding 1993J and 2013df, thus, we can only compare the previous one with the early-time LC of SN~2011dh. In the first $\sim$100 days, mid-IR LCs of SNe 1993J and 2011dh seem to be quite similar to each other, while 1993J produces a steeper decline later. It's also worth noting that, as described in detail in \citet{szalai16}, 2011dh has lower peak magnitudes in optical bands, but also show redder optical colors; this two effects can result the similar 3.6 $\mu$m evolution in the early phases. After the first $\sim$100 days, the main reason behind the different mid-IR LC evolution of the objects can be the various levels of extra IR radiation source (newly-formed dust, IR echo), which is also assumed by \citet{helou13} in the case of 2011dh.

A significant difference between the slopes of early (up to $\sim$250 days) and late-time LCs of SN~1993J can be noticed (2 mag(100d)$^{-1}$ vs. 0.03 mag(100d)$^{-1}$). This may indicate single or multiple event(s) between $\sim$250 and $\sim$3800 days that could result an incline in mid-IR, similarly to SN~2013df after $\sim$450 days \citep{maeda15,kamble16}. As \citet{maeda15} directly showed, late-time spectral signs of CSM interaction in SN~1993J and SN~2013df look quite similar (compared +626d and +670d spectra of SNe~2013df and 1993J, respectively). Moreover, \citet{smith17} showed that optical spectra of SN~1993J has not changed significantly between $\sim$2450 and $\sim$8350 days, which indicate a long-lasting continuous CSM interaction in its environment (see also later in Sections \ref{subsec:col} and \ref{subsec:rad}).

As an illustration, we also plotted the very late-time 4.5 $\mu$m LC of SN~1993J in Fig. \ref{fig:irlc_long}, which shows a steeper decline (0.05 mag(100d)$^{-1}$) than 3.6 $\mu$m does (just as can be spotted also in Fig. \ref{fig:lc}). While it is unusual  that dusty/interacting SNe become bluer at such late times \citep[see e.g.][]{szalai19}, we assume that here we see rather an observational effect (the object seems to fade into the local background after $\sim$7500 days on 3.6 $\mu$m images, thus, we can not follow its real decline steepness after that).

\subsection{Analitical SED models}\label{subsec:models}

To obtain the physical properties of the dust, we calculated the mid-IR spectral energy distributions (SEDs) of SN~1993J from the {\it Spitzer} data.

\begin{figure*}
	\includegraphics[width=\columnwidth]{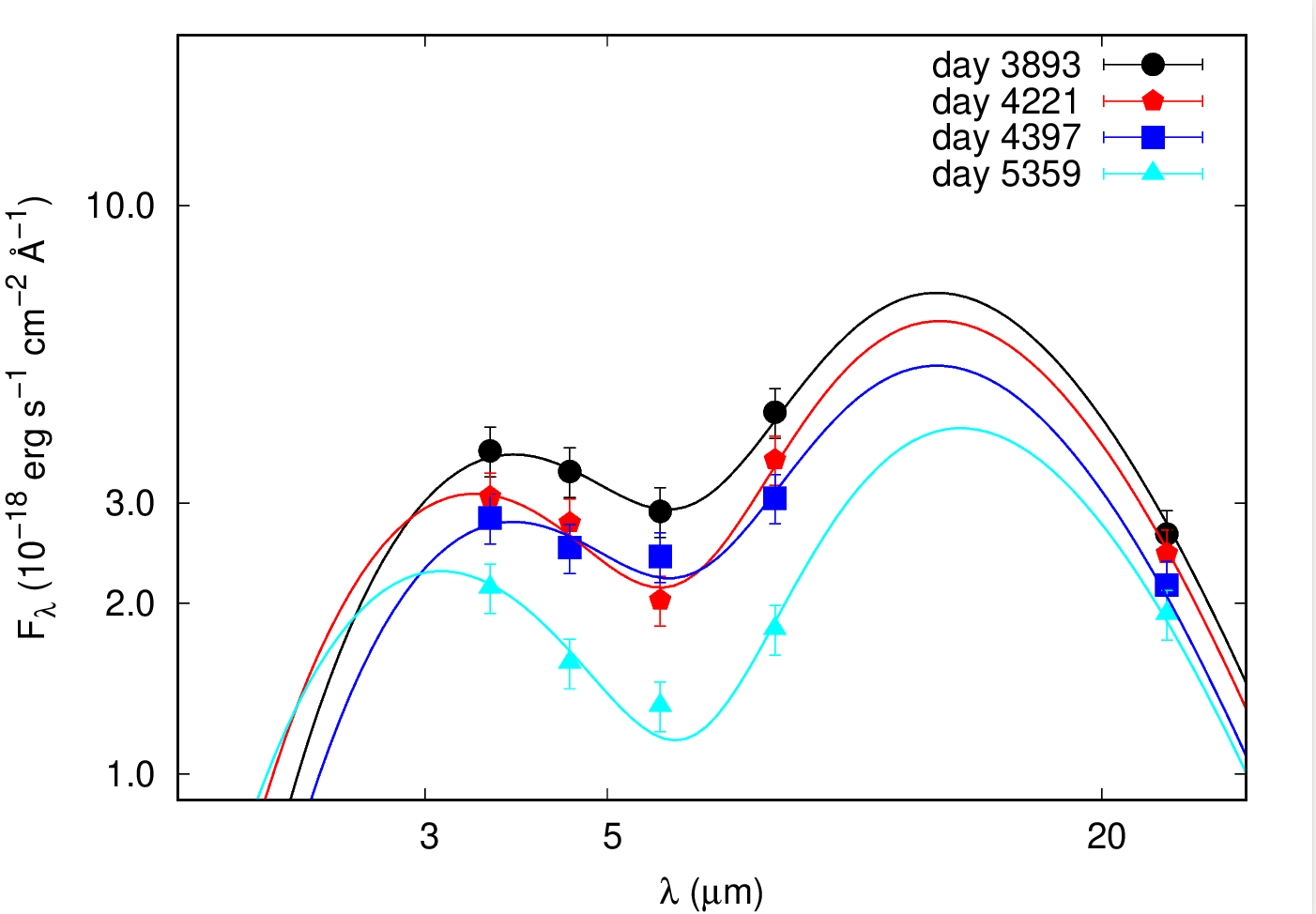}
		\includegraphics[width=\columnwidth]{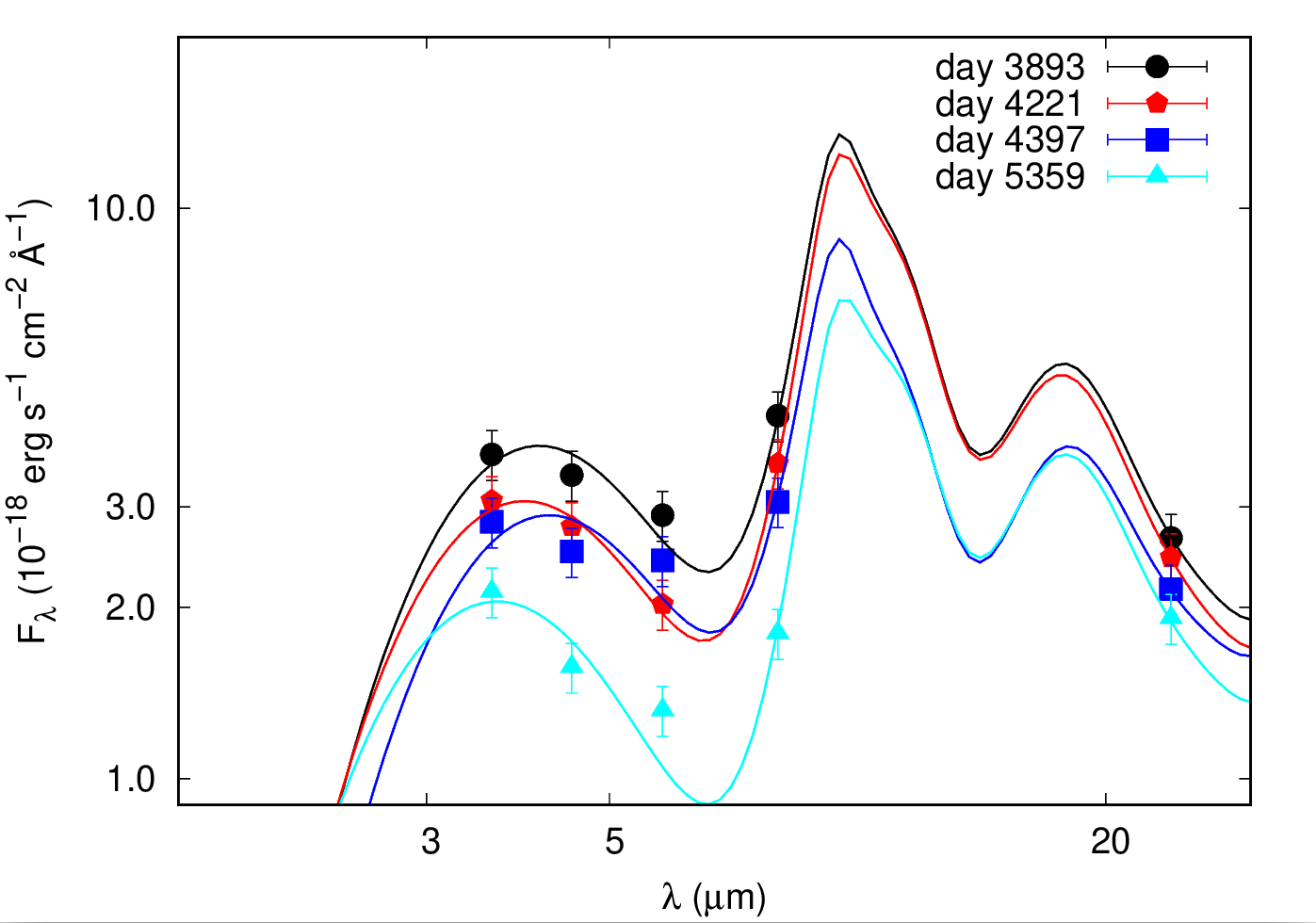}
    \caption{The best-fit two-component amorphous carbon ({\it left}) and C-Si-PAH ({\it right}) dust models on mid-IR SEDs of SN~1993J. Parameters of these models are listed in Table~\ref{tab:amc_si}.}
    \label{fig:amc_si}
\end{figure*}

First, it is necessary to check whether the late-time observed mid-IR flux can be explained only by the thermal radiation of the cooling ejecta. 
Since the mid-IR SEDs show a slow but unambiguous evolution (see in Fig.~\ref{fig:amc_si}), at least part of the observed late-time mid-IR flux probably connects to SN~1993J and its close environment.
As discussed e.g. by \citet{fransson_kozma02} for the case of SN~1987A, after $\sim$3000 days, $^{44}$Ti becomes the dominant source of the inner SN ejecta with a luminosity of L $\sim$ (10$^{36}-10^{37})$ erg s$^{-1}$; it can be used as a general estimation also for other CCSNe, see e.g. \cite{tanaka12}.
In a previous paper, \citet{zhang04} already showed that late-time integrated optical luminosity (2.85 $\times$ 10$^{38}$ erg s$^{-1}$, calculated from the latest known VRI data obtained at day 3245) is much above this level.
If we integrate the 3.6-24.0 $\mu$m fluxes at the first epoch of our {\it Spitzer} dataset (3893 days), we get an even larger value of  $L_\rmn{IR} \sim$1.15 $\times$ 10$^{39}$ erg s$^{-1}$. 
These findings are in accordance with the several signs of ongoing late-time CSM interaction in SN~1993J, while the high level of IR excess also indicate that (either newly-formed of pre-existing) dust grains in the environment of the SN probably give significant contribution to its observed luminosity.

In the next step, we fit both blackbody and analytical dust models to the mid-IR SEDs of SN~1993J. 
We only used epochs between 3875$-$5359 days when all the four IRAC channels and MIPS 24 $\mu$m data are available.
By combining the IRAC and MIPS data, we were able to constrain better fits and distinguish between the presumed multiple dust components.
Since SN~1993J shows a slow evolution (see in Fig.~\ref{fig:lc}), we were able to select three epochs of MIPS observations (days 3893, 4398, 5359), to which we could match IRAC observations obtained within a month. 
Following a similar strategy, we could also generate a fourth SED (fixed its epoch to 4221 days) interpolating the nearest IRAC and MIPS fluxes.
We adopted the dust models from \cite{hildebrand83} \citep[also applied by e.g.][]{doty94,fox10,fox11,sugerman12,szalai19} assuming both amorphous carbon and partly-silicate dust composition \citep[a mixture of C-Si-PAH,][called hereafter silicate dust]{weingartner01}. 
The widely used model assumes only the thermal emission of the optically thin dust with mass $M_d$, a particle radius $a$ at a single equilibrium temperature $T_d$ and at a distance $d$. The flux of the dust can be written as modified blackbody emission:
\begin{equation}
\label{eq:F}
    F_{\lambda}=\frac{M_d B_{\lambda}(T_d) \kappa(a)}{d^2}
\end{equation}
\noindent where $B_{\lambda}(T_d)$ is the Planck function and $\kappa_{\lambda}$ is the dust mass absorption coefficient as a function of the grain size:
\begin{equation}
\kappa_{\lambda}(a)=\left (\frac{3}{4\pi \rho a^3}\right)(\pi a^2 Q_v(a))
\end{equation}
\noindent where $\rho$ is the bulk density and $Q$ is the emission efficiency. 
We fit $T_d$ and $M_d$ as free parameters and adopted $\kappa_{\lambda}$ values from \cite{colangeli95} and \citet{weingartner01} for amorphous carbon and silicate models, respectively.
None of the examined mid-IR SEDs can be fitted by one-component (either blackbody or dust) models; thus, we used two-component models in the final step of the modelling, see Fig.~\ref{fig:amc_si} and Table~\ref{tab:amc_si} for the best-fit dust models and their parameters.

\begin{table*}
	\caption{Parameters for the best-fit two-component amorphous carbon and silicate dust models of SN~1993J.}
	\label{tab:amc_si}
	\begin{tabular}{l|ccc|ccc} 
		\hline
		\multicolumn{1}{c}{}
		& \multicolumn{3}{c}{Warm dust component}
		& \multicolumn{3}{c}{Cold dust component}\\
		Epoch & $M_\rmn{warm}$ & $T_\rmn{warm}$ & $L_\rmn{IR,warm}$ & $M_\rmn{cold}$ & $T_\rmn{cold}$  & $L_\rmn{IR,cold}$ \\
		(days) & $(10^{-3}M_{\odot})$ & (K) & $10^{39}$(erg/s) & $(10^{-3}M_{\odot})$ & (K) & $10^{39}$(erg/s)\\
		\hline
		\multicolumn{7}{c}{Amorphous carbon dust composition}\\
		\hline
    	3893 & 0.0007 & 640 & 0.29 & 1.64 & 190 & 1.56\\
		4221 & 0.0003 & 710 & 0.22 & 1.48 & 190 & 1.41\\
		4398 & 0.0005 & 640 & 0.22 & 1.22 & 190 & 1.16\\
		5359 & 0.0001 & 780 & 0.15 & 1.33 & 180 & 0.95\\
		\hline
		\multicolumn{7}{c}{Silicate dust composition}\\
		\hline
		3893 & 0.0170 & 550 & 0.64 & 4.72 & 180 & 1.27\\
		4221 & 0.0097 & 580 & 0.45 & 3.66 & 190 & 1.31\\
		4398 & 0.0163 & 530 & 0.53 & 5.80 & 160 & 0.82\\
		5359 & 0.0035 & 640 & 0.24 & 3.46 & 180 & 0.93\\
		\hline
	\end{tabular}
\end{table*}

\begin{figure}
	\includegraphics[width=\columnwidth]{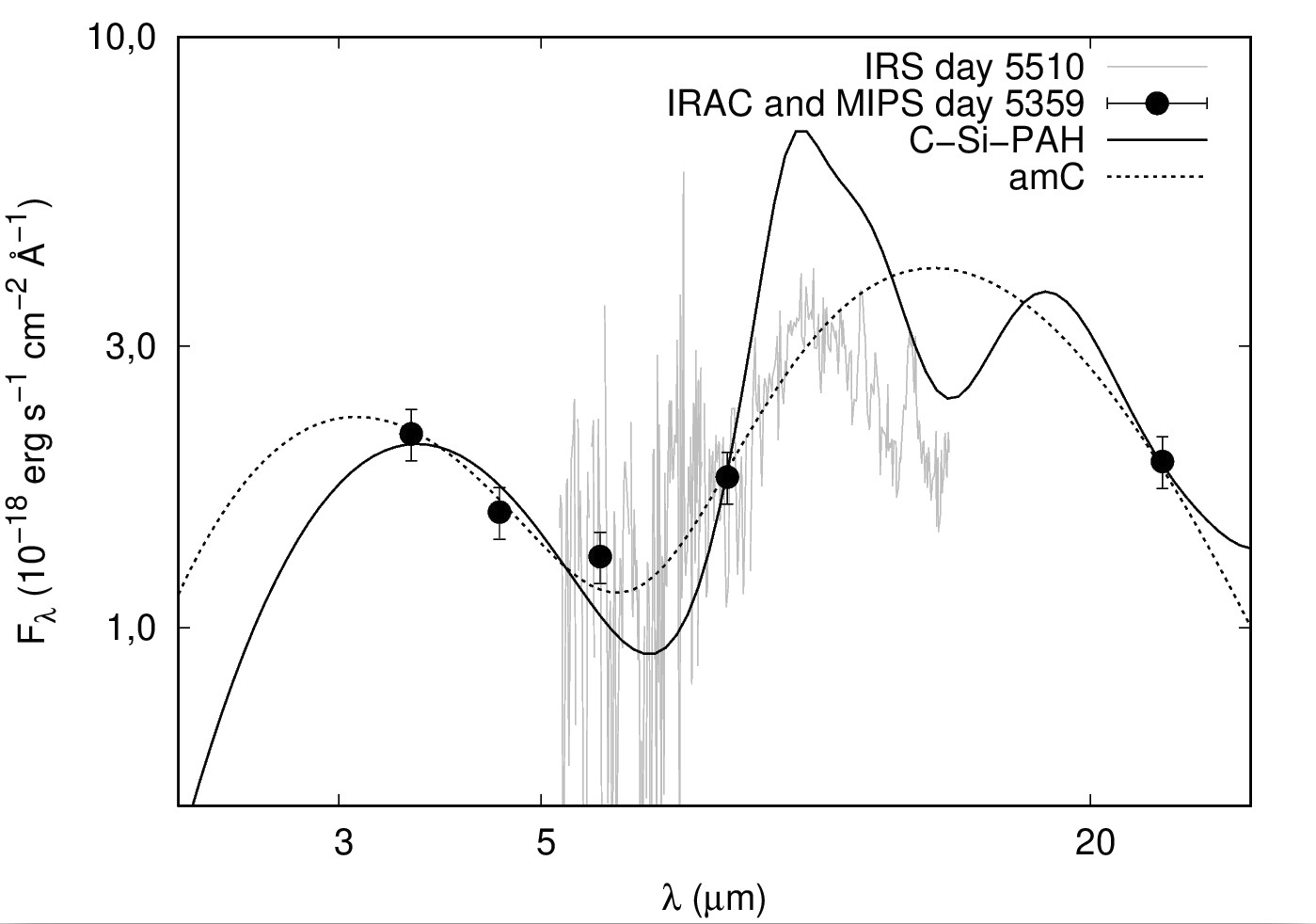}
    \caption{IRS spectrum of SN~1993J at day 5510 post-explosion, together with the best-fit two-component amorphous carbon (with dotted line) and C-Si-PAH (with solid line) models. The latter one seems to follow the spectrum quite well, supporting the assumption of presence of Si dust.}
    \label{fig:spectra}
\end{figure}

The total dust masses are $\sim(1.2-1.6)\times$ $10^{-3} M_{\odot}$ in case of amorphous carbon dust composition, and $\sim3$ times larger in case of silicate dust.
Taking a look at only the parameters of warm dust component, we can see that amorphous carbon models result two order of magnitude smaller dust masses and $\sim120$K larger temperatures, while the parameters of the cold component are similar assuming either amorphous carbon or silicate dust.

While either pure carbonaceous or silicate models give a good fit to the observed mid-IR SEDs, the single IRS spectrum of SN~1993J (obtained at day 5510, close to the epoch of the last complete SED) offers a good chance to distinguish between the two compositions.
Although the spectrum has a moderate signal-to-noise ratio, it can be seen well that our best-fit silicate model follows the IRS spectrum closely (see Fig.~\ref{fig:spectra}), supporting our assumption on the presence of silicate-based dust in SN~1993J.
Henceforward, we continue our analysis using exclusively the silicate dust models.

Considering that the temperatures of the warm and the cold component are fairly different, we can imagine them as two detached, thin dust shells. 
Similar conclusions were made before during the analyses of the few dust-forming SNe with complete mid-IR (3.6$-$24 $\mu$m) SEDs \citep[see e.g.][]{kotak09,fabbri11,szalai11,meikle11,szalai13}, and also in the case of SN~1987A \citep[which has a very extended dataset from optical to sub-mm wavelengths, see e.g.][and references therein]{matsuura19}.

Finally, it is worth pointing out that the model from our fittings based on the assumption that the observed IR flux originates entirely from optically thin dust. 
However, this may be generally not the case and may result in an underestimation of dust mass, since the fitted temperature represents a mass-weighted average dust temperature \citep[see][]{priestley20}. 
As a consequence, in an optically thick case, the analytical models do not describe the expanding remnant completely. 
Therefore, the optical depth of the assumed dust shells in the studied mid-IR range ($\tau_\rmn{shell}$) can characterize the fitted models. Following \citet{fox10}, we can check at least the self-consistency of our models, writing the optical depth as:
\begin{equation}
    \label{eq:tau}
    \tau_\rmn{shell}=\frac{M_d}{4\pi r^2} \kappa_\rmn{avg},
\end{equation}

\noindent giving originally by \citet{lucy89}. $\kappa_\rmn{avg}$=1000 {cm}$^2$ {g}$^{-1}$ is the average $\kappa$ value from our best-fit silicate model between 3.6 and 24 $\mu$m, considering single size dust grains.
Taking average dust masses of  $\approx$0.12$\times$10$^{-3} M_{\odot}$ and $\approx$4.41$\times$10$^{-3} M_{\odot}$ and average minimum (blackbody) radii of $R_\rmn{BB} \approx$0.41$\times$10$^{16}$cm and $\approx$2.89$\times$10$^{16}$cm (see Section \ref{subsec:size}) for the warm and the cold component, respectively, the equation yields $\tau_\rmn{shell} <$ 1.13 and 0.84; thus, we get back optically thin dust shells according to the original assumptions.

\subsection{Size estimations and specific radii}\label{subsec:size}
The location of the presumed dust shells plays a vital role in order to distinguish among the potential sources of dust grains. Newly-formed dust grains can be located in the inner (unshocked) ejecta as well as in the CDS between outer (shocked) ejecta and shocked CSM. 
On the other hand, pre-existing dust located beyond the vaporization radius can be heated collisionally (up to the edge of the forward shock) or radiatively (by the peak SN luminosity and/or by energetic photons arise from CSM interaction). 
For examining all these secenarios in detail, it is useful to give the fiducial radii and estimate the characteristic sizes of the expanding remnant.

As we hint above, blackbody radius ($R_\rmn{BB}$) serves as the minimum size of an optically thin dust shell \citep[see e.g.][]{fox10, fox11}.
We fit two-component blackbody models to the calculated mid-IR SEDs presented above. For the warm and the cold component, the blackbody radii are in a range of (0.05-0.10)$\times$10$^{16}$cm and (2.84-2.99)$\times$10$^{16}$cm, respectively.

The sizes of the ejecta and the shock front play an important role in differentiating among the possible sources. 
Fortunately, we have direct information about the radio size of the SN.
\cite{martividal11} presents  Very-Long-Baseline-Interferometry (VLBI) observations from day 60 to 4606 post-explosion from which the radio size of the SN is (3.57-3.80)$\times$10$^{17}$ cm at day 4606.
The standard model states that the radio emission comes from the region in between the reverse and forward shocks \citep{chevalier82a}, thus, radio measurements can be directly used for giving the size of the CDS.

One the other hand, we can estimate also the size of the ejecta from its velocity.
Based on the work of \cite{chevalier82a}, \cite{bjornsson15} use a formula of expansion velocity $v_\rmn{ej}=$1.6$\times$10$^4 (\rmn{t}/300\rmn{days})^{-0.2}\rmn{km s^{-1}}$ for the case of SN~1993J; from that, the presumable size of the ejecta is between 3.7$\times$10$^{17}$ cm at day 4606.
This is in agreement with the results of \cite{fransson05} who determined $\sim$7000km s$^{-1}$ as inner and $\sim$10\,000km s$^{-1}$ as outer velocity of the ejecta from an optical spectrum obtained at $\sim$ 1000 days, resulting 2.8$\times$10$^{17}$ cm and 4.0$\times$10$^{17}$ cm for the inner and outer radius of the ejecta at day 4606, respectively.
However, we note that the expansion of the ejecta decelerates after $\sim$3100 days \citep{weiler07,martividal11}, thus, the size of the SN is probably a bit smaller than the calculated values above.

\cite{bjornsson15} argued that the radio emission in SN~1993J is determined by the extent of the Rayleigh-Taylor instability arising from the contact discontinuity between these shocks. 
Thus, in case of a thin dust shell (the thickness of the dense shell formed by the shocked CSM and the SN ejecta is smaller than its radius), we approximate the size of the shell with values calculated from radio observations \citep{marcaide09,mvidal11}. 
As it is also summarized e.g. in \cite{bjornsson15} and \cite{kundu19}, the radio emitting shell can be assumed to be spherical and both its inner and outer radius follow the same time dependence of R$\propto t^m$ with m $\approx$(0.8-0.9) (where $t$ is the time after explosion and $m$ is the deceleration parameter) during the first $\sim$3100 days.
As a result, we can estimate the size of the shell with values calculated from radio observations.

If we assume pre-existing dust grains distributed in a shell with a light-crossing time greater than the duration of the optical emission, an IR echo is likely to form and heats the dust grains radiatively \citep{fox10}. 
To describe the IR echo properly, we must outline the geometry of the close SN environment. 
After the SN explosion, the peak SN luminosity disrupt most of the pre-existing dust and creates a dust-free cavity defined by the vaporization radius. 
Therefore, the IR echo is characterized by the radius of the dust-free cavity and of the external dust shell \citep{dwek85}.
The size of the vaporization radius ($R_\rmn{evap}$) is given by \citet{dwek85} \citep[also applied e.g. by][]{fox11}:

\begin{equation}
\label{eq:evap}
    R_\rmn{vap} = \left( \frac{L_\rmn{SN}}{16 \pi \sigma T_\rmn{v}^4 \langle Q \rangle} \right)^{1/2},
\end{equation}

\noindent where $L_\rmn{SN}$ is the peak UV-visual luminosity of the SN, $\sigma$ is the Stefan-Boltzmann constant, $\langle Q \rangle$ is the Planck-averaged value of the dust emissivity.
Since both our best-fit models and the available spectra support the silicate dust composition, we adopted $T_\rmn{v}=$1500K as the vaporization temperature of silicate grains \citep[][]{nozawa03} and $L_\rmn{SN} = 10^{42.5}\rmn{erg/s}$ \citep[][]{richmond94, shigeyama94}.
Thus, for $\langle Q \rangle =1$ (blackbody case), Eq. \ref{eq:evap} yields $R_\rmn{vap}$= 1.48$\times$ 10$^{16}\rmn{cm}$ for the radius of the dust-free cavity. 

In addition, as the IR echo is formed by travel-time effects, the duration of the echo ($t_\rmn{ech}$) defines the echo radius \citep[$R_\rmn{ech}$, see e.g. in][]{fox10,fox11}:
\begin{equation}
\label{eq:echo}
    R_\rmn{ech}=\frac{c t_\rmn{ech}}{2}.
\end{equation}

After clarifying the characteristic radii of the SN regions and of the presumable dust shell(s), we investigate the possible origin and heating mechanisms of dust in Sections~\ref{subsec:newly}-\ref{subsec:rad}.

\subsection{Newly-formed dust scenario}\label{subsec:newly}

In this part of our analysis, we investigate the viability of newly-formed dust scenario in SN~1993J.

Since minimal radii of presumed dust-containing regions calculated from blackbody fitting of mid-IR SEDs ($R_\rmn{BB} \approx$ (0.05-0.10)$\times$10$^{16}$ cm and (2.84-2.99)$\times$10$^{16}$ cm for warm and cold component, respectively) are much smaller then the estimated radii based on radio/optical measurements (see Section~\ref{subsec:size}), both dust components may entirely locate within the potential dust-forming regions of SN~1993J (practically, the unshocked ejecta and the CDS across the contact discontinuity, respectively). 
Nevertheless, to make conclusions on the real possibility of presence of newly-formed dust in SN~1993J, we also review our results in the context of theoretical expectations and of previous observational results.

Generally, type IIb SNe are not considered to be cosmic dust factories. Based on theoretical calculations, their small-mass hydrogen envelopes basically allow to form very small ($<<$0.01 $\mu$m) grains, which are thought to be destroyed by the propagation
of the reverse shock into the dust-forming region on a century-long timescale \citep[see for a review e.g.][]{kozasa09,gall11}. 
Nevertheless, as was presented in detail by \citet{nozawa10}, a significant amount (0.01-0.1 $M_{\odot}$) of dust can form in the ejecta of a Type IIb SN within a few years after explosion, in agreement with the observed dust content of the $\sim$350 years-old galactic remnant Cassiopeia A (Cas A) assumed to originate from a 93J-like explosion \citep[see e.g.][]{arendt99,arendt14,barlow10,delooze17}.

Regarding SN~1993J, fortunately, there is a possibility to directly compare the results of our analysis of late-time mid-IR data to that of an independent method carried out by \citet{bevan17}.
Analyzing a very late-time optical spectrum (obtained in 2009, relatively close to the epoch of our last mid-IR SED), \cite{bevan17} carried out an estimation of dust parameters via modeling [O III] $\lambda\lambda$ 5007,4959 {\AA} and [O II] $\lambda\lambda$ 7319,7330 {\AA} emission lines using their {\sc damocles} Monte Carlo line transfer code \citep{bevan16}. Their basic assumption was that the red–blue line-profile asymmetries and the presence of extended red emission wings can be explained by absorption and scattering effects of ejecta dust grains on the emitted photons \citep[as described first by][]{lucy89}.  

\citet{bevan17} used both smooth and clumped models with different grain composition (amorphous carbon and silicate) and grain sizes.
Their smooth models resulted dust masses in a range of (5$\times$10$^{-3}-0.12)M_{\odot}$, while using clumped dust distribution increased their calculated dust masses by $\sim$1.5 times. 
The authors concluded to prefer the Si dust models with the possible range of total dust mass of 0.08-0.15 $M_{\odot}$. 
While these values are larger than the ones we found during our mid-IR SED fittings ($\sim$5$\times$10$^{-3} M_{\odot}$), there are some important notes that may allow to get these results closer to each other. 
First, applying a clumpy model would also increase our dust masses even with an order of magnitude \citep[see e.g.][]{meikle07, priestley20}. 
Second, as an even more important factor, {\it Spitzer} data are not sensitive to very cold (T $\lesssim$ 50K) dust, in contrast with the method used by \citet{bevan17}. 
Such very cold local dust can be present in a large amount ($\sim$0.1-1 $M_{\odot}$) even in a decade-old SN ejecta, as was directly proven in the case of SN~1987A via far-IR and sub-mm data analysis \citep[see][and references therein]{matsuura19} and which can be also seen in the case of Cas A (see above). 

We also note that in the model of \cite{bevan17}, the dust forming shell has an internal radius of 2.4$\times$10$^{17}$cm and an outer radius of 3.2$\times$10$^{17}$cm, which is also in accordance with the results of radio and optical size measurements discussed above, strengthening the concept of newly-formed dust in the environment of SN~1993J.

Taking all of these into account, we can conclude that analysis of late-time mid-IR data supports the scenario of local dust formation in SN~1993J assumed by previous theoretical and observational studies. We can imagine the two mid-IR SED components as two detached dust shells: the warm component may locate in the unshocked (inner) ejecta, while the cold component could be slightly beyond the shocked ejecta, in the region of the CDS, just as concluded before in the cases of other dust-forming SNe \citep[see e.g.][]{kotak09,fabbri11,szalai11,meikle11,szalai13}.

\subsection{Collisional heating of pre-existing dust grains}\label{subsec:col}

Previously, we found that the observed late-time mid-IR excess can be explained by the  radiation of newly-formed dust locate in the ejecta/CDS. 
However, various direct signs of ongoing CSM-interaction have been observed in the vicinity of  SN~1993J (see Section~\ref{sec:intro}), which argue against the exclusiveness of the newly-formed dust and advocate the presence of pre-existing dust. 
\cite{smith17} found that flat-topped and double-peaked emission line profiles -- revealed first by \cite{matheson00} as direct indications of ongoing CSM interaction -- are still observable more than 20 years post-explosion \citep[connection of late-time spectra and ongoing CSM interaction in SN 1993J was also studied e.g. by][]{zhang04,fransson05,milisavljevic12}.

In this section, we examine whether pre-existing dust grains can be heated by the post-shocked gas due to the collision of the SN ejecta and the CSM \citep[see][and references therein]{fox10}. 
Following the method presented in \cite{fox10} \citep[and also applied by e.g. ][]{fox11,tinyanont16,szalai19}, we estimated the mass of dust processed by the forward shock of the SN. \citet{fox10} assume that the hot, post-shocked gas heats the dust shell and the total gas mass can be constrained from the volume of the emitting shell using the equations describing grain sputtering.
Assuming a dust-to-gas mass ratio of 0.01, the mass of dust is:

\begin{equation}
\label{eq:one}
M_{\mathrm{d}}(M_{\odot}) \approx 0.0028 \left( \frac{v_{\mathrm{s}}}{15\,000 \,\mathrm{km} \,\mathrm{s}^{-1}} \right)^3 \left( \frac{t}{\mathrm{year}} \right)^2 \left( \frac{a}{\mathrm{\mu m}}\right),
\end{equation}

\noindent where $v_\mathrm{s}$ is the shock velocity, $t$ is the time post explosion, and $a$ is the dust grain size. Note, that the derived formula discloses that the mass of the shocked gas is independent of the grain density. Based on the cited works, we used $v_\mathrm{s}$ = 5000 km s$^{-1}$ and 15\,000 km s$^{-1}$ for the lower and upper limits of the shock velocities assumed to be constant, respectively. Regarding grain size, we used 0.005 $\mu$m and 0.1 $\mu$m as lower and upper limits, respectively. Based on these assumptions, we get that 15 years after explosion, which is the latest epoch used during our SED modeling, dust mass can be heated collisionally is between (10$^{-4}-$0.06) $M_{\odot}$. The upper end of this range is much larger then the masses we get from our best-fit dust models; however, these results do not exclude the possibility of the collisional heating of pre-existing grains.

To investigate the validity of this scenario more deeply, we follow the work of \cite{tinyanont19}, who applied the light-curve modelling method of \cite{moriya13} to the long-term mid-IR LC of the interacting Type Ib SN~2014C. A general and widely-used analytical model of the collision of the SN ejecta with a dense CSM is the standard circumstellar interaction model \citep[SCIM,][]{chevalier82a,chevalier82b,cf94,chevalier03}. \cite{moriya13} generalized a solution of the dense shell for the cases of non-steady mass loss and applied their model to the bolometric LCs of interacting SNe. 
As was found by \citet{tinyanont19} in the case of SN~2014C and can be also seen in the case of SN~1993J (see Section \ref{subsec:res_lc}), mid-IR components dominate the very late-time SEDs of SNe; thus, fitting of SCIM can be applied to the integrated mid-IR LC.

The model assumes that the density of the SN ejecta follows a double power-law profile ($\rho_\rmn{ej}\propto r^{-\delta}$ in the inner part and $\rho_\rmn{ej}\propto r^{-n}$ in the outer part), while the density of the CSM follows a power-law profile ($\rho_\rmn{CSM}=Dr^{-s}$) and the interaction is powered by the kinetic energy of the collision of the CSM and the dense shell. We produce the late-time LC from the mid-IR fluxes and fit Eq. \ref{eq:L}, given by \cite{moriya13} to the integrated mid-IR LC:

\begin{equation}
    L(t)=2\pi \epsilon Dr_\rmn{sh}(t)^{2-s}\left(\frac{(3-s)M_\rmn{ej}\left(\frac{2E_\rmn{ej}}{M_\rmn{ej}}\right)^{0.5}}{4\pi Dr_\rmn{sh}(t)^{3-s}+(3-s)M_\rmn{ej}}\right)^3 ,
	\label{eq:L}
\end{equation}

\noindent where $D$ and $s$ describes the density profile of the CSM, $r_\rmn{sh}(t)$ is the radius of the expanding dense shell at the given epoch, $M_\rmn{ej}$ and $E_\rmn{ej}$ are the mass and the energy of the SN, while $\epsilon=0.1$ is the conversion efficiency from kinetic energy to radiation \citep[adopted its value also from][]{moriya13}.

First, to eliminate the degeneracy of the explosion parameters, we adopted them from previous studies and only fit the parameters of the CSM. We used the values of $M_\rmn{ej}$ and $E_\rmn{ej}$ determined from early-time bolometric LC fittings of SN~1993J \citep{nagy16}. We assume that the ejecta mass does not change significantly during the studied period. However, the kinetic energy of the expanding ejecta at these late epochs must be lower than it was at the time of the explosion. To estimate the kinetic energy of the shocked ejecta, we use an average value of ejecta velocity ($v_\rmn{ej}$=9260 km s$^{-1}$) calculated from the formula $v_\rmn{ej}=$1.6$\times$10$^4 (\rmn{t}/300\rmn{days})^{-0.2}\rmn{km s^{-1}}$ given by \cite{bjornsson15}; as a result, we get $E_\rmn{ej}=$1.84 $\times$10$^{50}\rmn{erg}$.

Second, we are able to determine the size of the shell with values calculated from radio observations \citep{marcaide09}. Although, the $r_\rmn{sh}(t)$ generally solved numerically, it can also be determined directly from very-long-baseline interferometry (VLBI) and further multi-wavelength radio measurements of SN~1993J up to $\sim$4600 days \citep[see e.g.][]{bietenholz01,bietenholz03,bartel02,weiler07,marcaide09,martividal11,mvidal11}.
Since \cite{moriya13} assumes that the thickness of the dense shell formed by the shocked CSM and the SN ejecta is smaller than its radius, our assumption seems to be a valid approximation. 
Observations show that the radio emitting shell can be assumed to be spherical and both its inner and outer radius follow the same time dependence of R$\propto t^m$ with $m\approx$0.8-0.9 (where $t$ is the time after explosion and $m$ is the deceleration parameter) during the first $\sim$3100 days (summarized e.g. in \citet{bjornsson15} and \citet{kundu19}). This is consistent with the self-similar solutions ($n >$5) of the SCIM \citep{chevalier82b}. Thus, $m$ is determined by $n$ and $s$ in terms of the expression $m = (n - 3)/(n - s)$.

During our fittings, we used six different combinations of the variables $s$ (1.66, 2) and $n$ (7, 10, 12) using characteristic values. $s=$2 refers to the general assumption of steady case mass-loss of the progenitor (its usability for SN~1993J has been also supported by e.g. \citet{bjornsson15}), while $s \approx$1.66 is suggested by studies based on radio \citep{marcaide97,marcaide09} and X-ray \citep{fransson96} data. For $n$ values, 10 and 12 are the general assumptions of \citet{moriya13} for SN Ib/c and RSG progenitors, respectively \citep[and also used before in the case of SN~1993J, see e.g.][]{baron95,marcaide09}, while $n \approx$7 has been suggested by \citet{bjornsson15}. All of these $n$ and $s$ combinations result values of $m$ in a range of 0.75-0.9, in accordance with radio measurements.

\begin{figure}
	\includegraphics[width=\columnwidth]{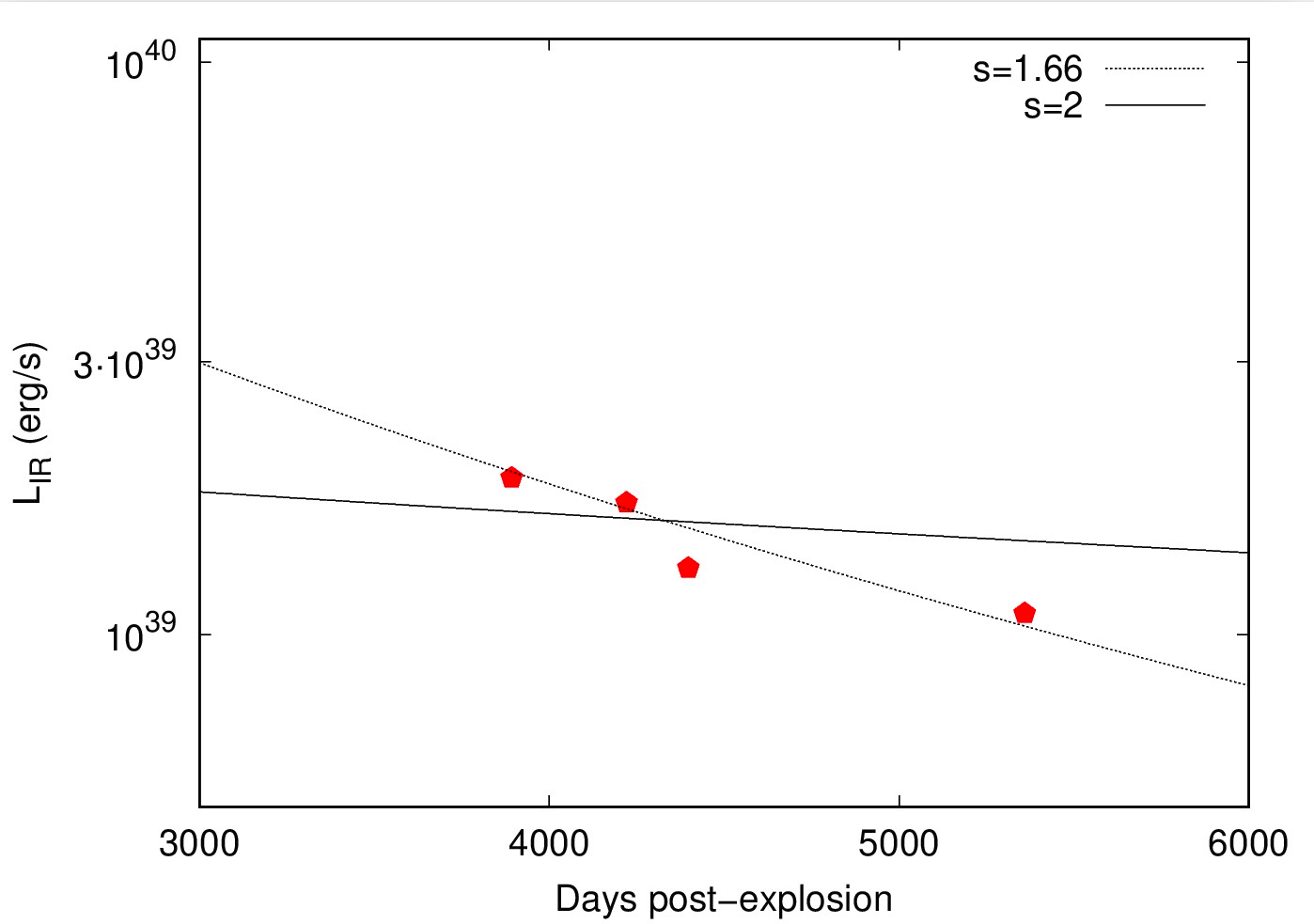}
    \caption{{Integrated mid-IR luminosity curves of SN 1993J constructed from the best-fit silicate dust models. We assumed that the density of the CSM follows a power-law profile ($\rho_\rmn{CSM}=D ^{-s}$) and during the fit we used two values of s: $s=2$ refers to the steady mass loss scenario which is plotted with solid line, while $s=1.66$ derived from radio and X-ray \citep{marcaide97, marcaide09} data \citep{fransson96} is plotted with dashed line.}}
    \label{fig:Lmod}
\end{figure}

Thus, the only fitted parameter was $D$, derived to be $D=10^{14.21}$ and 10$^{9.39}$ (cgs) for $s$=2 and 1.66, respectively (varying the values of $n$ does not change the values of $D$ significantly). We show two examples of the fit -- ($n$=7, $s$=2); ($n$=10, $s$=1.66) -- in Fig.~\ref{fig:Lmod}.
In the case of $s=2$, value of $D$ can be easily used for calculating the pre-explosion mass-loss rate via the formula of:
\begin{equation}
    D = \dot{M} / 4 \pi v_w,
\end{equation}
\noindent where, assuming wind velocity to be $v_w$ = 10 km s$^{-1}$ \citep[adopted from][]{fransson96}, we get  $\dot{M} \approx$ 3.27 $\times$10$^{-5} M_{\odot}$ yr$^{-1}$ for the pre-explosion mass-loss rate of the progenitor, which is in alignment with the results of \cite{fransson96}. Moreover, this value seems to be in a good agreement with $\dot{M} \approx(10^{-5}-10^{-4})M_{\odot}$ yr$^{-1}$ found via analysis of X-ray and radio data \citep[see e.g.][]{fransson96,fransson05,immler01,chandra09}. 

Nevertheless, if we also calculate the transition time ($t_t$) parameter given by \citet{moriya13} to check the validity of applying Eq~\ref{eq:L}, we get extremely large time scales (>10$^3$ days).
In addition, as pointed out e.g. by \citet{bjornsson15}, the break seen in both radio \citep[e.g.][]{weiler07} and X-ray \citep{chandra09} LCs at $t \approx$3100 days may indicate that the reverse shock is entering a roughly flat portion of the density distribution of the ejecta, leading to a drop in the intensity level of CSM interaction, as well as making the application of self-similar solutions questionable at these late-time epochs. 

Taking into account these last factors, it is not likely that collisional heating of pre-existing dust grains is a viable explanation for the late-time mid-IR excess observed in SN~1993J (however, some contribution cannot be ruled out). Instead, the permanent presence of CSM interaction seen e.g. in very-late time optical spectra gives strong motivation to also examine the radiative heating scenario (see next Section).

\subsection{Radiative heating of pre-existing dust}\label{subsec:rad}

While, as noted above, its level seems to decrease after $\sim$3100 days, very-late time optical spectra \citep[$\sim$2450$-$8350 days,][]{matheson00,smith17} indicate a continuous presence of SN-CSM interaction in the environment of SN~1993J.
It seems to be also in accordance with the -- unfortunately missing, but presumably plateau-like -- period of the long-term mid-IR LC of the SN between $\sim$250 and $\sim$3800 days (see Fig.~\ref{fig:irlc_long}).
Both datasets may imply a constant heating source, for which radiative heating scenario of distant pre-existing dust grains may be a plausible explanation.

To  examine the viability of the radiative heating mechanism more deeply, we adopted a simple IR echo model from \cite{bode80} and \cite{dwek83} \citep[also used by e.g.][]{fox10,fox11}. 
In this model, a central point source heats a spherically symmetric shell of dust with a single radius and temperature and the emitting region is within this shell. According to the nature of the heating source and the geometry of the pre-existing dust, \cite{fox10} outline three main possibilities as follows \citep[also see in][]{fox11}. 
If the dust is distributed spherically symmetrically around the progenitor, the SN peak luminosity vaporizes some part of the dust and heats the remaining part of it closely to the vaporization temperature. As an alternative, if the dust is distributed in a shell at a radius larger than the vaporization radius, the SN peak luminosity heats the inside of this shell but only to the observed temperature. Lastly, if dust is distributed in a shell between vaporization and echo radii, the late-time optical luminosity -- generated by the CSM-interaction -- continuously heats the dust shell (this option is rather the reprocessing of the optical luminosity than a traditional IR echo).

Due to the incomplete data of SN~1993J, it is difficult to distinguish between the different IR echo scenarios. However, considering the first one, we can estimate the timescale of the possible plateau from the size of the dust-free cavity. The calculated vaporization radius are small to produce an IR echo on a decade-long timescale. Furthermore, the dust temperatures from our best-fit models (see in Table~\ref{tab:amc_si}) are significantly lower than the vaporization temperature for silicate dust \citep[$\sim$ 1500K,][]{nozawa03}. Thus, it is not likely that the SN peak luminosity could be a dominant heating source for pre-existing dust grains at late-times in case of SN~1993J. 

In the last scenario, the late-time optical luminosity, generated by the CSM interaction, heats continuously the dust shell, which is located between the vaporization and echo radii, also referred as the CSM interaction echo model \citep{gerardy02}. Taking into consideration the signs of ongoing CSM interaction in the vicinity of the SN (see above), this much more slowly changing emission seems to be a more possible heating source.

Since even the simplest IR echo models have numerous parameters and the available IR dataset of the SN~1993J is incomplete, we only estimate the size of the presumed dust shell in order to check the possibility of the radiative dust heating scenario. Assuming a steady state mass loss ($r^{-2}$ density distribution) and silicate composition, the mass of the shell can be written \citep[][]{dwek83}:

\begin{equation}
\label{eq:dustmass}
    M_s=4\pi \left ( \frac{4 \rho_{gr} a}{3 Q_v} \right ) \frac{\tau_d}{Z_d} \times \frac{R_1 R_2^2}{R_2 - R_1},
\end{equation}
\noindent where $R_1$ and $R_2$ is the inner and outer radius of the shell, $\tau_d=0.01$ is the UV-visual optical depth, $\rho_{gr}$=3 g cm$^{-3}$ is the mass density of grain, $Z_d=0.01$ \citep[adopted from][] {fox10} is the dust-to-gas ratio, and $a=0.1\mu m$ is the representative grain size in the shell. For our calculations, we adopted general values of dust parameters from \cite{dwek83}. We adopted the vaporization radius as the inner radius of the shell ($R_1$) and calculate the outer radii ($R_2$) from our best-fit silicate dust models. Thus, by solving the equation above, we get $R_2=$1.50$\times$ 10$^{16}$cm and $R_2=$1.16$\times$ 10$^{18}$cm for the outer radii of the dust shell. The first solution is very close to the vaporization radius and $\approx$ 2 times smaller than the average blackbody radius of the cold component, thus most likely does not describe an achievable shell geometry. The second solution gives a more conceivable outer radius of the dust shell, since its a few times larger both than the calculated ejecta and radio size of the SN at the epochs of the SEDs. To check the validity of the second solution, we calculated the duration of the presumable echo rising from the dust shell using Eq.~\ref{eq:echo}. This yields an $\sim$895 day-long plateau on the LC of the SN, which is an order of magnitude larger than the ones expected from the SN peak luminosity heating scenario. Hence, it strengthens the role of CSM interaction in pre-existing dust heating.

Finally, we also note here that \cite{sugerman02} revealed details of (at least) two optical light echo structures \citep[also examined by][]{liu03} in the wider environment of SN~1993J.
The echos are at 81 and 220 pc (2.50$\times$10$^{20}$cm and 6.79$\times$10$^{20}$ cm, respectively) in front of the SN and their material is consistent with the HI surface density detected in this region of the host galaxy, indicating that they are presumably interstellar dust structures illuminated by the SN. 
Well, considering that the {\it Spitzer}/IRAC apertures cover only a $\sim$1.96$\times$10$^{20}$cm radius field on the sky and the SN is located in a well-isolated part of its galaxy, it is not likely that the previously detected interstellar echos contribute significantly to the flux detected on late-time {\it Spitzer} images.

\section{Discussion and conclusions}\label{sec:concl}

We carried out an extensive mid-IR analysis of SN 1993J from 10 up to 26 years of its evolution, based on a mainly unpublished archival dataset of Spitzer Space Telescope. We present long-term LCs at 3.6, 4.5, 5.8, 8.0, and 24.0 $\mu$m, and SEDs at 4 epochs (between 3893 and 5359 days) calculated from these mid-IR fluxes (see in Fig.~\ref{fig:lc}). 
After ascertained that the cooling ejecta cannot be the only source of the observed late-time mid-IR fluxes, we examined the origin of this mid-IR excess in SN~1993J and considered both newly-formed and pre-existing heated dust scenarios. 

To reveal the physical properties of the presumed dust in the environment of SN~1993J, we fitted both blackbodies and analytic (pure amorphous carbon and silicate-based) dust models to the late-time mid-IR SEDs of the SN. We found that these SEDs can be only modeled with two-component (warm and cold) dust, similarly to that of another few SNe, for which detailed mid-IR analysis was possible to carry out before. Taking a rare opportunity, an IRS spectrum close to the last SED allowed us to find evidence for a (partially) silicate-based dust composition (see in Fig.~\ref{fig:spectra}), which we considered throughout the latter parts of our analysis.

Due to the several model parameters and the lack of data, it is difficult to describe the presumable dust entirely and accurately. Although, special parameters such as the location and the mass of the dust grains carry vital information about the origin as well as the dust forming processes. 

Our best-fit silicate models give dust masses in a range of (3.5-6.0)$\times$10$^{-3}M_{\odot}$ (see in Table~\ref{tab:amc_si}).
The fitted SED models describe the thermal emission of dust without reference to its origin; thus, for investigating the possibility of ongoing local dust formation, we compared our conclusions with that of theoretical expectations and with the results of a late-time optical spectral analysis by \cite{bevan17}, see in Section~\ref{subsec:newly}.
The latter authors examined the integrated and indirect effects of dust grains on emitted optical photons of expanding ejecta, while our analysis helps to reveal dust temperature, and thus, to disentangle warm and (not-so-) cold dust components. 
The dust masses presented by \cite{bevan17} are larger than the values from our best-fit models, however, they can be comparable taking into account the potential effect of dust clumping as well as the presence of very cold ($<$50K) local dust.
It is also worth remarking that the model from our fittings may result an underestimation of dust mass anyway, since it is based on the assumption that the observed IR flux originates entirely from optically thin dust \citep[][]{priestley20}. 

We also connected the late-time {\it Spitzer} data of SN~1993J with its early-time L-band photometry \citep{matthews02}. 
There is only a slight difference between the 3.6 $\mu$m/L fluxes measured at $\sim$250 and at $\sim$3800 days, which could be explained either by a quasi-continuous plateau or by the decay of single/multiple event(s) that can result in mid-IR re-brightening phase(s) (similarly to the case of SN~2013df, the spectroscopic twin of SN~1993J, between $\sim$450-800 days).
\citet{matthews02} explained the early-time near-IR fluxes with the thermal emission of dust with a temperature of $T_\rmn{dust} \sim 850$K and a mass of $M_\rmn{dust} \sim 10^{-5} M_{\odot}$. 
These are in a good agreement with the parameters of the warm dust component we determined during the analysis of late-time {\it Spitzer} data (at the same time, because of the lack of longer-wavelength data, we have no information on the presence of the cold dust component before $\sim$3800 days). 

If we assume that we see newly-formed (warm) dust at both early and later epochs, it raises several questions.
Applying again the formula for $v_\rmn{ej}$ by \citet[][see Section \ref{subsec:col}]{bjornsson15}, it results $\sim$3.58$\times$10$^{16}$cm for the ejecta radius on day 250, which is approximately the same as the blackbody radius calculated from late-time {\it Spitzer} data. 
It could mean that we see the same warm dust at $\sim$250 and after $\sim$3800 days. However, it would require an extremely inefficient cooling of the grains (which, taking into account the low probability of the presence of large grains in a Type IIb ejecta, is less probable).
Another option is the continuous formation of ejecta dust. \cite{wesson15} found that in case of SN~1987A, the evolution of the dust mass can be described with a sigmoid function advocating that most of the dust formed in later times (90 per cent of the dust would form between $\sim$3000-14\,000 days post-explosion). It means that even the dust formed at $\sim$250 days cooled or destroyed up to $\sim$3800 days, it could be replaced by newer dust formed during at a single or at multiple events at later phases.
While it can be a viable option, we cannot really prove it, since there are no direct observational evidence for that during the mid-IR LC gap. As we also showed it in Section \ref{subsec:res_lc}, we cannot really compare long-term mid-IR evolution of SN~1993J to that of other SNe IIb (or, in fact, to that of any other SNe except the very unusual case of SN~1987A) to find any analogues.

At the same time, as we noted above, several signs indicate a long-term, continuous presence of SN-CSM interaction in the environment of SN~1993J 
\citep[e.g. the quasi-constant late-time optical spectra, see][]{matheson00,smith17}, which can be also connected with a long-lasting mid-IR plateau via heating of pre-existing dust grains.
As we showed in Sections \ref{subsec:col} and \ref{subsec:rad}, neither collisional heating model, nor heating of grains by SN peak luminosity seem to play a dominant role in the late-time mid-IR radiation of SN~1993J. Instead of them, radiative heating of pre-existing grains by energetic photons emerging from ongoing CSM interaction is probably a viable option.

As a summary, we came to the conclusion that the observed late-time mid-IR excess of SN~1993J may be explained by either dust-forming processes (assuming newly-formed dust both in the unshocked ejecta and in the cold dense shell, just as seen in Cas A remnant and also assumed in other dust-forming SNe in a few years after explosion), or heating of pre-existing dust grains via ongoing CSM interaction. 
Yet, on account of the available incomplete data and some uncertainties throughout the analysis, it is difficult to exactly determine the dust parameters and to give their exact origin.
At the same time, we showed that a long-term mid-IR dataset allows us to distinguish between the possible formation and heating processes of dust grains in SN environments.

\section*{Acknowledgements}
{We thank our anonymous referee for valuable comments.}

This project has been supported by the GINOP-2-3-2-15-2016-00033 project of the National Research, Development and Innovation Office of Hungary (NKFIH) funded by the European Union, and by NKFIH/OTKA FK-134432 grant.
T.S. is supported by the J\'anos Bolyai Research Scholarship of the Hungarian Academy of Sciences. T.S. and S.Z. is supported by the New National Excellence Program (UNKP-20-5, UNKP-20-3, and UNKP-21-3) of the Ministry for Innovation and Technology from the source of the National Research, Development and Innovation Fund. A.P.N is supported by the NKFIH/OTKA PD-134434 grant.

%%%%%%%%%%%%%%%%%%%%%%%%%%%%%%%%%%%%%%%%%%%%%%%%%%
\section{Data Availability}

The data underlying this article are available in Spither Heritage Archive and in Combined Atlas of Sources with Spitzer IRS Spectra. The datasets were derived from sources in the public domain: 

\noindent \href{https://sha.ipac.caltech.edu/applications/Spitzer/SHA/}{https://sha.ipac.caltech.edu/applications/Spitzer/SHA/}

\noindent \href{https://cassis.sirtf.com/}{https://cassis.sirtf.com/}


\bibliographystyle{mnras}
\bibliography{main.bib}

%%%%%%%%%%%%%%%%%%%%%%%%%%%%%%%%%%%%%%%%%%%%%%%%%%

%%%%%%%%%%%%%%%%% APPENDICES %%%%%%%%%%%%%%%%%%%%%

%\appendix


%\newpage

%%%%%%%%%%%%%%%%%%%%%%%%%%%%%%%%%%%%%%%%%%%%%%%%%%

\bsp	
\label{lastpage}
\end{document}